# Tien Shan data on inelastic proton-air cross section at 10 PeV

**N. Nesterova**
P N Lebedev Physical Institute RAS.

New data on the absolute value on the inelastic proton–air cross section at 10 PeV are presented. Conclusion is made about the growth with energy of the inelastic proton–air cross section according to comparisons with experimental data were obtained at the Tien Shan complex array on various components of extensive air showers: hadrons, Cherenkov light and electrons with many different calculated models of cosmic rays interactions at the atmosphere. The analysis showed that the rise conforms to 7-9 % per one order of the energy from 0.2 TeV (accelerators with fixed targets) to 10 PeV (EAS cosmic rays.). That corresponds to around 350 mb at 1 PeV and 380 mb at 10 PeV of primary cosmic rays. These data correspond better to the QGSJET-II-04 model version.

**Keywords:** Inelastic p–air cross-section from 0.2 TeV to 10 PeV. Extensive air showers.

## 1. Introduction

The purpose of this work was to find the law of the rise of $\sigma_{p\text{-air}}$ - the inelastic proton–air cross section with the rise of the PCR energy. It was done on the base of the Tien Shan complex array [1] experimental data at primary cosmic rays (PCR) energies $E_0 = 0.5\text{-}10$ PeV and their extrapolation from 0.2 TeV ( experiments on accelerator with fixed targets). Experimental data of various components of extensive air showers (EAS) initiated by PCR were compared with many different former and modern simulation models. Our initial publications had been [2, 3]. Comparisons with some other current experimental data up to VHE are shown too.

## 2. Experimental results and comparisons with different models

The complex Tien Shan array (43.04 N, 76.93 E, P=685 g. cm$^{-2)}$) contained different EAS detectors: hadrons (the ionization calorimeter), electrons (scintillation and GM counters), muons (underground GM counters) and atmospheric Cherenkov light [4]. EAS were classified according to the total number of electrons $N_e$ ($N_e \sim E_0$) at the Tien Shan level.

Firstly Cherenkov light Q ( photon× m$^{-2}$) lateral distributions at R= 50 - 250 m from the axis of EAS at $E_0 = 1\text{-}10$ PeV PCR based on experiments at the Tien Shan [4] (Figure 1) and the former Pamir [5] arrays were compared with model calculations [4]. Experimental EAS "cascade curves" (Ne as a function of the atmosphere depth P (g cm$^{-2}$) at the constant EAS intensity I at $E_0 > 2$ PeV were received for comparing with calculations too [6]. In these models were taken various rise of the inelastic proton–air cross section $\sigma_{p\text{-air}}$: 0%, 7%, and 10% per one order of $E_0$ from $\sigma_0$ (0.2 TeV) = 265 mb (accelerator data). Conclusions were made from these experiments that $\sigma_{p\text{-air}}$ rise is ~ (7-9) % per one order of energy magnitude up to 10 PeV PCR.

The main conclusions were made on the base of the analysis of EAS hadron energy spectra at hadron energies $E_h > 1$ TeV of EAS in various intervals of electron number Ne ($E_0$). The special procedure of the processing for separation of each individual hadron in the ionization calorimeter and its control was done.

It was shown many simulations that the number of hadrons $N_h$ ($E_h$=1–5 TeV) at $E_0 = (0.5\text{–}10)$ PeV is practically independent of PCR mass composition, but it is sensitive to some interaction parameters, especially to the $\sigma_{h\text{-air}}$ and the inelasticity coefficient $K_{inel}$.

Initially experimental results on $N_h$ ($E_h > 1$ TeV, $N_e$) were compared [7] with former calculation models [8,9,10,11,12] for different $\sigma_{p\text{-air}}$ values at the coefficient at inelasticity $K_{inel}$ = var. Data on $N_h$ ($E_h > 1$ TeV, $N_e$) as well as our data on EAS Cherenkov light and $N_e$ "cascade curves" indicate that the rise of $\sigma_{p\text{-air}}$ is 7–9% and $\sigma_{p\text{-air}}$ (1PeV) = (350+15) mb, if the inelasticity coefficient $K_{inel}$ = 0.65±0.05 and $\sigma_{p\text{-air}}/\sigma_{\pi\text{-ir}}$=1.30±0.08, if $\sigma_{p\text{-air}} = \sigma_0$ (1+$\alpha$ ln $E_0$), $\sigma_0$ = 265 mb at 0.2 TeV.

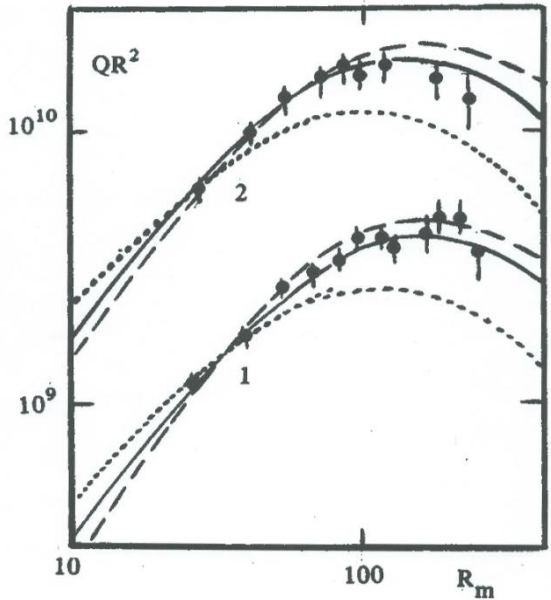
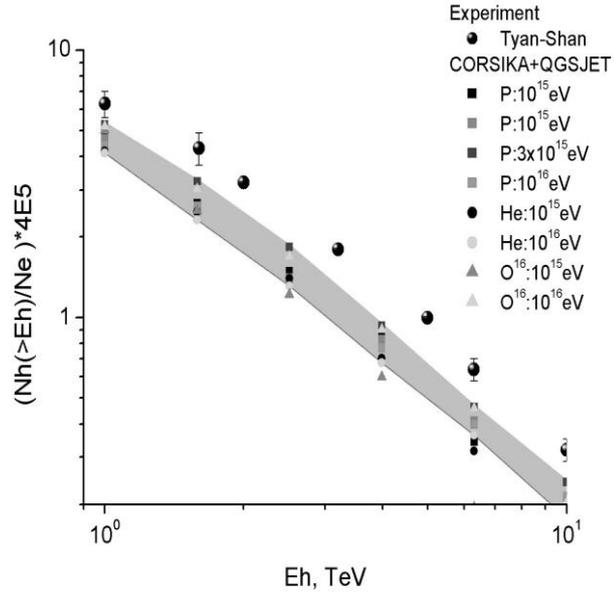

**Figure1.** Cherenkov light lateral distribution at Tien Shan. 1. $E_0$=2 PeV, 2. $E_0$=9 PeV. Dashed line - 0%, solid line - 7%, dotted line - 10% the rise of $\sigma$ inel. p-air is per one order of $E_0$.
**Experiment: black circles**.

Fig.2. Integral energy spectra at Tien Shan. The number of hadrons per shower (normalized to $N_e$=4 $10^5$ ($E_0 \approx$ 1 P V). For $E_0$=1, 3 and 10 PeV from primary: P (squares), He (circles), O (tranglets). Shaded area - QGSJET model.
**Experiment: black circles**.

After that we compared the experimental hadrons energy spectra with CORSIKA + QGSJET modern models with the same $N_e(E_0)$ intervals [13, 14, 15]. Values of $N_e(E_0)$ at the Tien Shan were received in the same special calculations.

Spectra for different primary nuclei (p, He, O) were examined by QGSJET- 0I model are sown in Fig.2. Integral energy spectra at Tien Shan for $E_0$=1 PeV, 3 PeV, 10 PeV PCR energies are presented. Calculations show that the number of hadrons per shower $N_h/N_{EAS}$, at $E_h$= (1 − 5) TeV is practically independent of the PCR mass composition.

So, the number of hadrons in the experiment exceeds the number in this version of the QGSJET model (Fig.2). Experimental hadrons energy spectra confirmed the correctness of our conclusions to 10 PeV PCR energies.

We compared experimental spectra with data of the QGSJET-II-3 model version for primary protons [14]. Results of the comparison with the QGSJET-II-3 indicated that the number of hadrons in the experiment outnumbers the simulated number too.

The new version of QGSJET-II model (QGSJET-II-04) was presented at 32nd ICRC [16]. Changes of model were based on analysis of recent LHC data on soft multi-particle production. In the new version the rise of $\sigma_{p-air}$ is about (8 − 9) % per one order of $E_0$ and $\sigma_{p-air}$ (1 PeV) ≈ 360 mb. This rise $\sigma_{p-air}$ is slower than in previous versions of QGSJET-II and better corresponds to our experimental data. Data of calculations (a copy from [16]) and our results as well as other experimental data of last years at $E_0 >$ 1 PeV are shown in Fig. 3.

## 3. Conclusion.
Our analysis based on Tien Shan experimental results on EAS of PCR at 0.5 – 10 PeV always shows a slow the rise of the cross section $\sigma_{p-air}$ with increasing energy. Conclusions are based on comparison of experimental data with different models on EAS hadron spectra as well as EAS Cherenkov light lateral distributions and "cascade curves", $N_e$ (P). This rise of the inelastic proton–air cross section corresponds to (7-9) % (not more than 10%) per one order of energy magnitude from 0.2 TeV (accelerators with fixed targets) to 10 PeV (EAS).

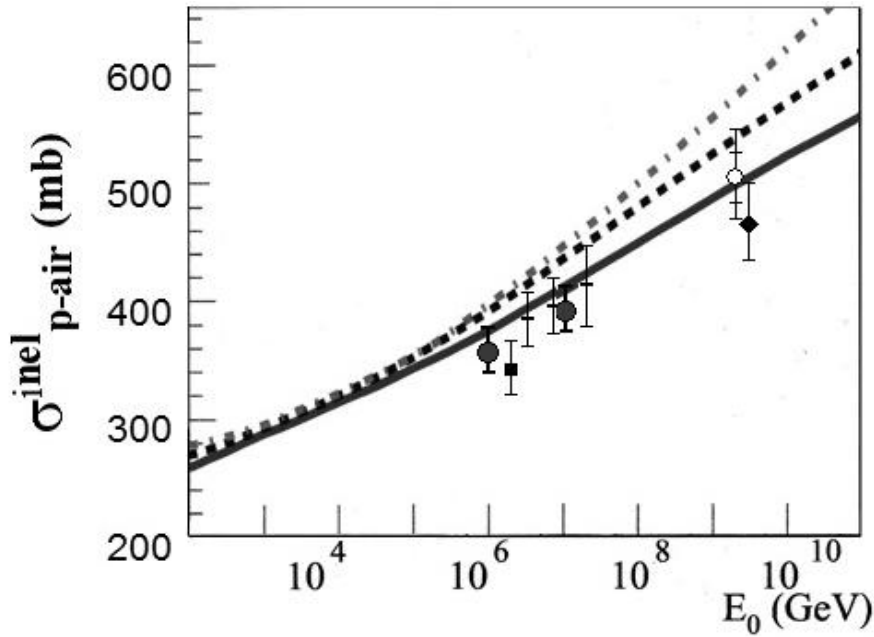

**Figure 3.** Proton-air cross section $\sigma_{p\text{-air}}$ vs PCR primary energy $E_0$ by the new version of QGSJET-II-4 (solid line), QGSJET-II-03 (dashed line), SIBYLL (dot-dashed line). Notation: Tien Shan data (two black circles) and some recent experimental data (HiRes, RAO, Yakutsk, EAS TOP).

The main conclusion made on the base of the method of analysis of EAS hadron energy spectra at $E_h > 1$ TeV. This value has an advantage that it is almost independent of mass composition of PCR at $E_0 = 0.5 - 10$ PeV in accordance with QGSJET models and other former models of calculations. This growth can proceed to $2 \cdot 10^{18}$ EeV according to Pierre Auger Observatory and HiRes data.

If conclusions based on experimental data are right, dissipation of the PCR energy in air is less than it is predicted by such models as CORSIKA+ QGSJET-0I, old QGSJET- II (-01, -02, -03), SIBYLL, and some previous models. In our recent works [14, 15, 16] the conclusion was made that it would be desirable to decrease $\sigma_{p\text{-air}}$ in QGSJET-0I and old QGSJET- II models. The new version of QGSJET-II model (QGSJET-II-04) [17] corresponds better to our and other recent data (HiRes, (RAO, Ulrich, et al. [18], Gamma, Knurenko et al. [19], EAS TOP, Aglietta et al. [20], ARGO-YBJ, Aielli at all. [21]).

**Acknowledgments.** Author thanks colleagues in the development and operation of the Tien Shan array and help in the process of the work and L.G. Sveshnikova for the QGSJET- calculation.